\documentclass[UTF-8,preprint,showpacs,preprintnumbers,superscriptaddress]{revtex4}
\pdfoutput=1
\usepackage{multirow}
\usepackage{diagbox}
\usepackage{booktabs}
\usepackage{array}
\usepackage{natbib}
\usepackage{amsmath,amssymb,graphicx,bm,subfigure,float,siunitx,color}
\usepackage{bm,subfigure,float,siunitx,graphicx}
\usepackage[utf8]{inputenc}
\usepackage[figuresright]{rotating}
\usepackage{color}
\usepackage{mathrsfs}

\begin{document}

\title{Possible evidences for physics beyond $\Lambda$CDM from DESI DR2 data}

\author{Rong-Jia Yang \footnote{Corresponding author}}
\email{yangrongjia@tsinghua.org.cn}
\affiliation{College of Physics Science and Technology, Hebei University, Baoding 071002, China}

\begin{abstract}
We analyze DESI DR2 data with a model-independent method and find that: (a) the expansion of the universe may speed up with a confidence level more than 2.3 $\sigma$ at redshift $z_{51}\in (0.51, 0.955)$; (b) the expansion of the universe may speed down with a confidence level greater than 1.7 $\sigma$ at redshift $z_{75}\in (0.955, 1.484)$; (c) $w_{\rm{x}}\leq w_{\rm{t}}<-1$ with confidence level exceeding 1.6 $\sigma$ at redshift $z_{53}\in (0.922, 0.955)$.
\end{abstract}

\maketitle

\section{Introduction}
Since the discovery of the phenomenon of accelerated expansion of the universe, $\Lambda$CDM (with an equation of state (EoS) $w_{\rm x}=p_{\rm x}/\rho_{\rm x}=-1$) has attracted widespread attentions for its simplest and most theoretically based approach. However, it also faces the crisis of cosmological constant problem \cite{Carroll:2000fy} and age problem \cite{Yang:2009ae,Vagnozzi:2021tjv}. In addition, the Hubble tension has posed new challenges for $\Lambda$CDM \cite{Riess:2019cxk}.

Recently, the Dark Energy Spectroscopic Instrument (DESI) collaboration presented a cosmological analysis based on the latest baryon acoustic oscillations (BAO) measurements from its Data Release 2 (DR2) \cite{DESI:2025zgx}. Their results point to discrepancies between datasets becoming more relevant within the $\Lambda$CDM model, preferring dynamical dark energy (DDE) as a possible solution \cite{DESI:2024mwx,DESI:2024hhd,DESI:2025zgx}. The DESI DR2 results are consistent with the $\Lambda$CDM model, but they exhibit a $2.3$ $\sigma$ tension \cite{DESI:2025zgx} with cosmic microwave background (CMB) measurements (Planck data including external CMB lensing data from \cite{ACT:2023dou}). Using the CPL parameterization, DESI DR2 data combined with CMB temperature and polarization anisotropies, as well as CMB lensing, shows
DDE is preferred at $3.1$ $\sigma$, increasing up to $4.2$ $\sigma$ when including SNe data \cite{DESI:2025zgx} (this preference is
also supported by the Dark Energy Survey BAO and SNe combined analysis \cite{DES:2025bxy}). It is important to note that Refs. \cite{DES:2024jxu,DES:2025bxy} report preference of a DDE without DESI data set. Other recent researches also suggest DDE, see, for example \cite{DESI:2025wyn,Chaussidon:2025npr,Teixeira:2025czm,DESI:2025gwf,
Scherer:2025esj,Specogna:2025guo,Liu:2025mub,Yang:2023qsz,Yang:2025kjq}. In \cite{Luongo:2024fww}, however, $w$CDM was favored by DESI 2024 data.

Using the Lagrange mean value theorem, a model-independent method without making assumptions about a specific DE model, a parameterized Hubble function, or a parameterized EoS for DE was proposed to analyze $H(z)$ parameter data, and the results suggest possible evidences for physics beyond $\Lambda$CDM \cite{Yang:2023qsz,Yang:2025kjq}. Here we generalize the method to analyze DESI DR2 data and examine whether we can obtain similar evidences beyond $\Lambda$CDM. We find that the universe may experience an accelerated expansion during $0.51<z<0.955$ and the EoS of DE may be less than -1 during $0.922<z<0.955$.

The paper is structured as follows. In Sec. II, we will present DESI DR2 data and the method needed to analyze the data. In Sec. III, we will provide the data obtained from the analysis and the results derived from them. In Sec. IV, we will give some conclusions and discussions.

\section{DESI DR2 data and methodology}
In this Section, we will outline 8 DESI DR2 data released recently and generalize the method proposed in \cite{Yang:2023qsz,Yang:2025kjq}, which is needed in data analysis.
\subsection{DESI DR2 data}
Baryon acoustic oscillations from galaxy surveys constrain the expansion history of the universe
during $0.1 < z < 4.2$, probing the matter dominated era and the recent era of cosmic acceleration.

The most precise BAO measurements to date come from the DESI, which incorporate observations of millions of galaxies and quasars. The measurements provide the unprecedented dataset which enables the construction of detailed 3D maps of the cosmic web, facilitates precise measurements of the universe's expansion
history \cite{DESI:2025zgx,DESI:2024mwx}, and provides robust constraints on the EoS parameter of DE \cite{DESI:2025fii,DESI:2024aqx}.

The BAO measurements of DESI are expressed as the transverse comoving distance $D_{\rm M}/r_{\rm d}$, the
angle-averaged distance $D_{\rm V}/r_{\rm d}$, and the Hubble horizon $D_{\rm H}/r_{\rm d}$, all normalized to
the comoving sound horizon at the drag epoch $r_{\rm d}$. Here we are interested in the $F$ data from DESI DR2 data \cite{DESI:2025zgx}, see the Table \ref{t1}, where
\begin{eqnarray}
\label{ap}
F\equiv D_{\rm M}/D_{\rm H}.
\end{eqnarray}
For a spatially flat universe the Hubble distance $D_{\rm H}$ and the transverse comoving distance $D_{\rm M}$ are defined, respectively, as
\begin{eqnarray}
\label{ha}
D_{\rm H}=\frac{1}{H(z)},
\end{eqnarray}
and
\begin{eqnarray}
\label{dm}
D_{\rm M}=(1+z)D_{\rm A}=\int_0^{z}\frac{1}{H(z')}dz',
\end{eqnarray}
where
\begin{eqnarray}
\label{da}
D_{\rm A}=\frac{1}{1+z}\int_0^{z}\frac{1}{H(z')}dz',
\end{eqnarray}
with $H(z)$ the Hubble function.

\begin{table*}
\begin{tabular}{l|l|l}
\hline
index & $z$ & $F$  \\
\hline
$z_1$ & 0.510 & $0.622\pm 0.017$   \\
\hline
$z_2$ & 0.706 & $0.892\pm 0.021$   \\
\hline
$z_3$ &  0.922 & $1.232\pm 0.021$   \\
\hline
$z_4$ &  0.934 & $1.223\pm 0.019$   \\
\hline
$z_5$ &  0.955 & $1.220\pm 0.033$   \\
\hline
$z_6$ &  1.321 & $1.948\pm 0.045$   \\
\hline
$z_7$ & 1.484 & $2.386\pm 0.136$   \\
\hline
 $z_8$ & 2.330 & $4.518\pm 0.097$   \\
\hline
\end{tabular}
\caption{$F$ data obtained from DESI's BAO measurement \cite{DESI:2025zgx}.}
\label{t1}
\end{table*}

\subsection{Methodology}
From the Planck 2018 results: $\Omega_{\rm K0}=0.001\pm 0.002$ \cite{Planck:2018vyg}, we consider a universe described by the spatially flat Friedmann-Robertson-Walker-Lema\^{i}tre (FRWL) metric
\begin{eqnarray}
\label{frwmet}
ds^2=-dt^2+a^2(t)\left[dr^2+r^2(d\theta^2+\sin^2\theta
d\phi^2)\right],
\end{eqnarray}
with $a(t)$ the scale factor. The Friedmann equations are given by
\begin{eqnarray}
&&H^2\equiv \left(\frac{\dot{a}}{a}\right)^2=\frac{8\pi G}{3}\rho, \\
\label{acc}
&&\frac{\ddot{a}}{a}=-\frac{4\pi G}{3}\left(\rho+3p\right),
\end{eqnarray}
where the dot indicates the derivative with respect to the $t$ and the unit $c=1$ is used. The total energy density $\rho$ and the corresponding pressure $p$ include contributions coming from the radiation, the nonrelativistic matter, and other components. To determine whether the expansion of the universe is speeded up, a physical quantity, called as the deceleration parameter, is required, which is defined as
\begin{eqnarray}
\label{de}
q=-\frac{a\ddot{a}}{\dot{a}^2}.
\end{eqnarray}
In data fitting, we often need another form of expression for it
\begin{eqnarray}
\label{q1}
q=-1+(1+z)\frac{1}{H}\frac{dH}{dz}=-1+(1+z)\frac{1}{E}\frac{dE}{dz}.
\end{eqnarray}
where $E(z)=H(z)/H_0$. In terms of the deceleration parameter, the total EoS, $w_{\rm t}=p_{\rm t}/\rho_{\rm t}$, can be written as
\begin{eqnarray}
\label{wt}
w_{\rm t}=-\frac{1}{3}+\frac{2}{3}q.
\end{eqnarray}
For a spatially flat universe, the parameter $F$ reduces to
\begin{equation}
F=E(z) \int_0^z \frac{1}{E(x)} d x,
\end{equation}
which implies
\begin{eqnarray}
\label{ed}
\frac{1}{E(z)}\frac{dE(z)}{dz}=\frac{1}{F}\left(\frac{dF}{dz}-1\right).
\end{eqnarray}
Therefore the deceleration parameter can be rephrased as
\begin{eqnarray}
\label{q3}
q=-1+\frac{1+z}{F}\left(\frac{dF}{dz}-1\right).
\end{eqnarray}

Now we generalize the mode-independent method introduced in \cite{Yang:2023qsz,Yang:2025kjq} to establish a basic framework for analyzing DESI DR2 data. Assuming that the parameter $F$ (therefore the Hubble function or the EoS of DE) is continuously differentiable, we have from Lagrange mean value theorem in Calculus
\begin{eqnarray}
\label{f1}
F^{\prime}(z_{ij})\equiv\frac{dF}{dz}\big{|}_{z=z_{ij}}=\frac{F(z_i)-F(z_j)}{z_i-z_j},
\end{eqnarray}
for any redshift interval with $z_j<z_{ij}<z_i$. If approximating the value of $F(z_i)$ at redshift $z_i$ with the datum $F_{\rm o}(z_i)$ at 1 $\sigma_{{\rm{F}}}(z_i)$ confidence level, then we can approximate $F^{\prime}(z_{ij})$ as
\begin{eqnarray}
\label{af}
F^{\prime}(z_{ij})\simeq \frac{F_{\rm o}(z_i)-F_{\rm o}(z_j)}{z_i-z_j},
\end{eqnarray}
at 1 $\sigma_{\rm{F'}}$ confidence level, where
\begin{eqnarray}
\label{ef}
\sigma_{\rm{F}^{\prime}}=\frac{\sqrt{\sigma^2_{\text{F}_{i}}+\sigma^2_{\text{F}_{j}}}}{z_i-z_j}.
\end{eqnarray}

Now considering the approximation of deceleration parameter in Eq. \eqref{q3} and using Eq. (\ref{f1}), we have
\begin{eqnarray}
\label{q4}
q(z_{ij})&=&-1+\frac{(1+z_{ij})}{F(z_{ij})}\left[\frac{dF}{dz}(z_{ij})-1\right]\nonumber,\\
&\simeq& -1+\frac{(2+z_i+z_j)}{F_{\text{o}}(z_{i})+F_{\text{o}}(z_{j})}\left[\frac{F_{\text{o}}(z_{i})-F_{\text{o}}(z_{j})}{z_i-z_j}-1\right],
\end{eqnarray}
at 1 $\sigma_{\rm{q}}$ confidence level, where $\sigma_{\rm{q}}$ is given by
\begin{eqnarray}
\label{eq}
\sigma_{\text{q}}=\frac{2+z_i+z_j}{(z_i-z_j)\left(F_{\text{o}i}+F_{\text{o}j}\right)^2}
\sqrt{\left[-2F_{\text{o}i}+(z_i-z_j)\right]^2\sigma^2_{\text{F}_j}+\left[2F_{\text{o}j}+(z_i-z_j)\right]^2\sigma^2_{\text{F}_i}}.
\end{eqnarray}
In Eq. (\ref{q4}), we have adopted the mid-value approximate method \cite{Yang:2023qsz}: $z_{ij}\simeq (z_i+z_j)/2$ and $F(z_{ij})\simeq [F(z_i)+F(z_j)]/2\simeq [F_{\rm{o}}(z_i)+F_{\rm{o}}(z_j)]/2$.

From Eqs. (\ref{q4}), (\ref{eq}), and therefore (\ref{wt}), we see that the observational systematics in $\sigma_{\text{F}_i}$ and the redshift binning $z_i-z_j$ will affect the reliability of our results. The larger $\sigma_{\text{F}_i}$ or the smaller $z_i-z_j$, the greater $\sigma_{\text{q}}$ will be.

\section{Applications}
When using Eqs. (\ref{q4}), (\ref{eq}), and (\ref{wt}) to analyze the observational $F$ parameter data from Table \ref{t1}, the error $\sigma_{\text{q}}$ would be amplified if $z_i-z_j\ll 1$. So we take the following limitations during the analysis process to make the results credible: $0.1\lesssim z_i-z_j\lesssim 0.5$. The obtained $q$ and $w_{\rm t}$ data are presented in Table \ref{t2} with 1 $\sigma$ confidence level are. From these data, we can draw the following conclusions:

(a) During the period $0.51< z< 0.955$, the universe may experience an accelerated phase, see for example, the expansion of the universe may accelerate with a confidence level exceeding 2 $\sigma$ at redshift $z_{41}\in (0.51, 0.934)$; a significance greater than 2.3 $\sigma$ at redshift $z_{51}\in (0.51, 0.955)$; more than 1.6 $\sigma$ at redshifts $z_{52}\in (0.706, 0.955)$ and $z_{53}\in (0.922, 0.955)$.

(b) During the period $0.934< z< 1.484$, the universe may experience a decelerated phase, see for example, the expansion of the universe may decelerate with a confidence level exceeding 1 $\sigma$ at redshifts $z_{64}\in (0.934, 1.321)$ and $z_{74}\in (0.934, 1.484)$; more than 1.6 $\sigma$ at redshift $z_{65}\in (0.955, 1.321)$; greater than 1.7 $\sigma$ at redshift $z_{75}\in (0.955, 1.484)$.

(c) We have $w_{\rm{x}}\leq w_{\rm{t}}< -1$ with a confidence level more than 1 $\sigma$ at redshift $z_{43}\in (0.922, 0.934)$; exceeding 1.6 $\sigma$ at redshift $z_{53}\in (0.922, 0.955)$. $w_{\rm{x}}< -1$ is also predicted from quintom model \cite{Yang:2024kdo} and CPL parameterization \cite{DESI:2024mwx}.

According to the Planck 2018 results \cite{Planck:2018vyg}, the phase transition from deceleration to acceleration of the universe occured at the redshift $z\simeq 0.632$ for a spatially-flat $\Lambda$CDM, which is consistent with the result (b) here, however, it is not consistent with the result (c) and the accelerated expansion of the universe at redshifts $z_{52}$ and $z_{53}$ in the result (a). The results (a) and (c) may suggest possible evidence for physics beyond $\Lambda$CDM. There are other evidences beyond $\Lambda$CDM come from interacting DE \cite{Pan:2019gop,Goswami:2025uih,Bansal:2025usn,Petri:2025swg}, early DE \cite{Sakstein:2019fmf,Poulin:2018cxd}, parameterized scenarios \cite{He:2024jku,
Paul:2025wix,DESI:2024mwx,DESI:2025wyn}, Hubble parameter data analysis \cite{Yang:2023qsz,Yang:2025kjq},  and modified gravity \cite{Yang:2024kdo,Ishak:2024jhs}.

\begin{table*}
\begin{tabular}{l|l|l}
\hline
index & $q$  & $w_{\rm t}$  \\
\hline
$z_{41}\in (0.51, 0.934)$ & $-0.22\pm 0.11$ & $-0.48\pm 0.07$ \\
\hline
$z_{51}\in (0.51, 0.955)$ & $-0.35\pm 0.15$ & $-0.57\pm 0.10$ \\
\hline
$z_{42}\in (0.706, 0.934)$ & $-0.22\pm 0.22$ & $-0.48\pm 0.14$ \\
\hline
$z_{52}\in (0.706, 0.955)$ & $-0.45\pm 0.27$ & $-0.63\pm 0.18$  \\
\hline
$z_{43}\in (0.922, 0.934)$ & $-3.75\pm 3.70$ & $-2.83\pm 2.47$  \\
\hline
$z_{53}\in (0.922, 0.955)$ & $-3.16\pm 1.89$ & $-2.44\pm 1.26$  \\
\hline
$z_{64}\in (0.934, 1.321)$ & $0.17\pm 0.16$ & $-0.22\pm 0.10$  \\
\hline
$z_{74}\in (0.934, 1.484)$ & $0.36\pm 0.26$ &   \\
\hline
$z_{65}\in (0.955, 1.321)$ & $0.33\pm 0.20$ & \\
\hline
$z_{75}\in (0.955, 1.484)$ & $0.48\pm 0.28$ &  \\
\hline
\end{tabular}
\caption{$q$ and $w_{\rm t}$ obtained from DESI's BAO measurement.}
\label{t2}
\end{table*}

\section{Conclusions and discussions}
Generalizing the model-independent method proposed in \cite{Yang:2023qsz,Yang:2025kjq}, we have analyzed DESI DR2 data with the corresponding confidence levels. From the obtained data, we come to the following conclusions:

(a) the universe may have an accelerated phase during the era $0.51< z< 0.955$;

(b) the universe may have a decelerated phase during the era $0.934< z< 1.321$;

(c) The EoS of DE may be less than $-1$ during the era $0.922<z<0.955$.

The result (c) suggests that there may exist accelerated phase before the current accelerating period.

The obtained $q(z)$ and $w_{\rm t}$ data may be used to test cosmological models. The obtained results here depend on the DESI DR2 data. In future researches, more and more accurately measured BAO data are needed to validate our results. The method provided here could be generalized to analyze other observational data.

\begin{acknowledgments}
This study is supported in part by National Natural Science Foundation of China (Grant No. 12333008) and Hebei Provincial Natural Science Foundation of China (Grant No. A2021201034).
\end{acknowledgments}

\textbf{Appendix}
Assuming that $x_1$, $x_2$, ..., $x_n$, with uncertainties $\delta x_1$, $\delta x_2$, ..., $\delta x_n$, the uncertainty for the measurement $f_{\rm m}$ of $f(x_1, x_2, ..., x_n)$ is given by
\begin{eqnarray}
\delta f=\sqrt{\left(\frac{\partial f}{\partial x_1}\delta x_1\right)+...+\left(\frac{\partial f}{\partial x_n}\delta x_n\right)}.
\end{eqnarray}
We say that the measurement is at $(|f_{\rm m}|/\delta f)$ $\sigma$ confidence level.

\bibliographystyle{ieeetr}
\bibliography{H1}
\end{document}